# Physics-driven universal twin-image removal network for digital in-line holographic microscopy

**Mikołaj Rogalski,[a,*] Piotr Arcab,[a] Luiza Stanaszek,[b] Vicente Micó,[c] Chao Zuo,[d,e,f] Maciej Trusiak[a,**]**

[a]Institute of Micromechanics and Photonics, Warsaw University of Technology, 8 Sw. A. Boboli St., 02-525 Warsaw, Poland
[b]NeuroRepair Department, Mossakowski Medical Research Institute, Polish Academy of Sciences, 5 A. Pawlińskiego St., 02-106 Warsaw, Poland
[c]Departamento de Óptica y Optometría y Ciencias de la Visión, Universidad de Valencia, C/Doctor Moliner 50, Burjassot 46100, Spain
[d]Smart Computational Imaging Laboratory (SCILab), School of Electronic and Optical Engineering, Nanjing University of Science and Technology, 210094, Nanjing, Jiangsu Province, China
[e]Smart Computational Imaging Research Institute (SCIRI) of Nanjing University of Science and Technology, 210019, Nanjing, Jiangsu Province, China
[f]Jiangsu Key Laboratory of Spectral Imaging & Intelligent Sense, 210094, Nanjing, Jiangsu Province, China

**Abstract**. Digital in-line holographic microscopy (DIHM) enables efficient and cost-effective computational quantitative phase imaging with a large field of view, making it valuable for studying cell motility, migration, and bio-microfluidics. However, the quality of DIHM reconstructions is compromised by twin-image noise, posing a significant challenge. Conventional methods for mitigating this noise involve complex hardware setups or time-consuming algorithms with often limited effectiveness. In this work, we propose UTIRnet, a deep learning solution for fast, robust, and universally applicable twin-image suppression, trained exclusively on numerically generated datasets. The availability of open-source UTIRnet codes facilitates its implementation in various DIHM systems without the need for extensive experimental training data. Notably, our network ensures the consistency of reconstruction results with input holograms, imparting a physics-based foundation and enhancing reliability compared to conventional deep learning approaches. Experimental verification was conducted among others on live neural glial cell culture migration sensing, which is crucial for neurodegenerative disease research.

*Mikołaj Rogalski**, E-mail: mikolaj.rogalski.dokt@pw.edu.pl
**Maciej Trusiak**, E-mail: maciej.trusiak@pw.edu.pl

## 1 Introduction

One of the main challenges of optical microscopy is the problem of observing transparent samples, e.g., biological cells, which due to the very limited absorption are only slightly visible with generally very low contrast impeding examination and diagnostics. This issue may be bypassed by the quantitative phase imaging[1] (QPI) techniques, which allow not only to increase the imaging contrast, but also provide quantitative information about the optical thickness of the sample. QPI techniques may be divided into two groups: (1) interferometric (e.g., interferometry[2], holography[3]) and (2) intensity (e.g., Fourier ptychography[4], transport of intensity equation[5]) methods. Interferometric techniques provide a well-defined access to complex field via interference (fringe) pattern analysis, whereas intensity techniques computationally estimate phase information from intensity-only images without employing the interference phenomenon. Among the vast family of holographic techniques, Gabor principle[6] digital in-line holographic microscopy[7,8] (DIHM) seems especially interesting because of its simplicity and cost-effectiveness. However, contrary to other interference-based techniques, it does not provide access to phase part of complex optical field at the detector plane.

The most straightforward DIHM configuration, so-called lensless DIHM, is composed of only three elements: a coherent light source, a sample and a camera, Fig. 1. Implemented in two opposite layouts[9] lensless microscopes can provide extremely large field of view (FOV) imaging (limited by the camera sensor size) without magnification and modest resolution limits (limited by camera pixel size)[10] or limited FOV with higher magnifications (ranging from 5X to 20X)[11]. To increase the imaging resolution (at a cost of decreasing FOV), there may be employed DIHM configurations with microscope objectives[12], that also may be applied in a simple and low-cost manner[13]. DIHM principle of operation bases on measuring weakly scattering samples. When such object is illuminated, small part of the light is scattered and greater part of the light passes unaffected. At the camera plane the result of interference between those two wavefronts forming an in-line hologram is observed.

The conventional approach for reconstructing quantitative phase information in DIHM involves the numerical backpropagation[14] of a hologram's intensity distribution to the focal plane of the sample (Z distance). However, due to the absence of phase information of the complex optical field at the camera plane in DIHM, the backpropagation of a phase-free hologram to the object plane results in the superposition of the object's optical field with its digitally generated twin, defocused at the minus Z distance. As a consequence, the reconstructed in-focus complex optical field exhibits characteristic double-defocused Gabor fringes, commonly known as the twin-image effect, Fig. 1. Up to this day, researchers actively studied this central problem and proposed several solutions to this pivotal issue. The most popular is the Gerchberg-Saxton (GS) algorithm[15,16], which requires to collect several (at least 2) different holograms (acquired with different wavelengths[17,18] or Z distances[19]) and then to iteratively propagate the complex-field between the hologram planes to retrieve phase factor of the complex field in hologram plane with appropriate recorded intensity constrains. The main disadvantage of this solution is that efficiently collecting several different holograms significantly complicates the system (hardware and software-wise) – also from an economical point of view, elongates the measurement time, and therefore eliminates one of the greatest advantages of DIHM – its simplicity and robustness. When it comes to a single hologram twin-image removal, then again GS algorithm may be used, but this time hologram is propagated between object and hologram planes employing appropriate constraints in the object plane[20,21]. Nevertheless, such solution requires vast a priori knowledge about sample (often to mask object regions) and usually is not as effective as multi-hologram GS as it works without data-multiplexing. Interesting group of solutions includes the regularization approaches[22,23], where at the object plane twin-image is iteratively diffused employing various norms (e.g., total variation), whereas object sharp features remain unchanged. However, again this requires some a priori information about the measured object and usually necessitates a large number of iterations, making such solutions a time-consuming one.

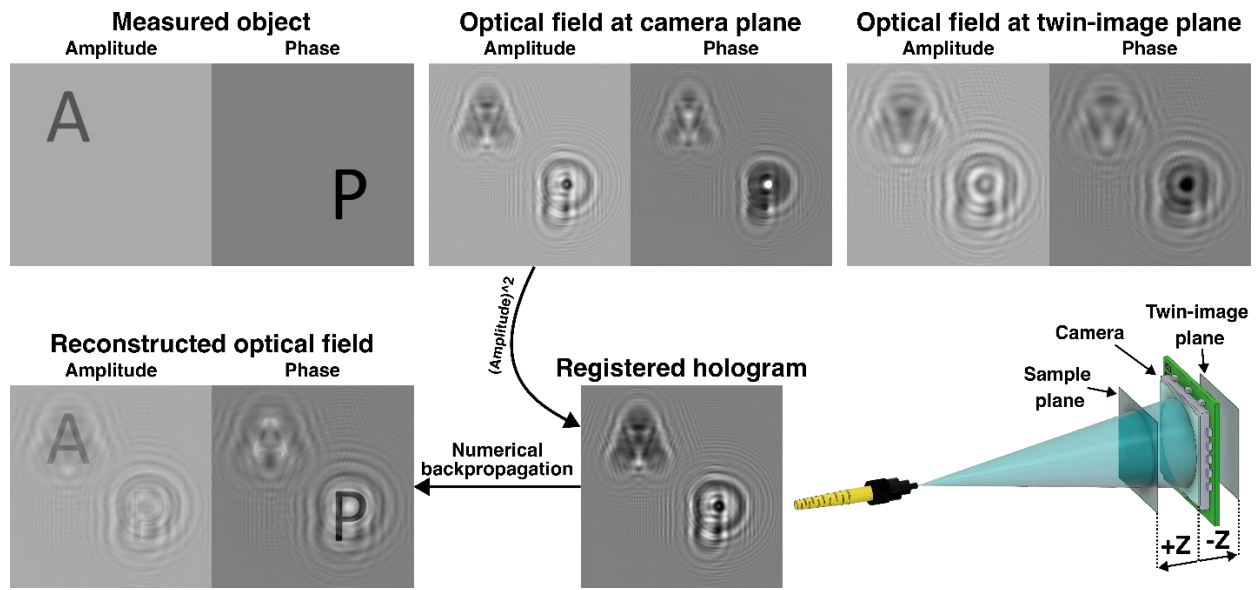

**Fig. 1** Scheme of a typical lensless DIHM system, simulation of optical fields at different planes and exemplary reconstruction of the hologram with the use of numerical backpropagation. Amplitude images are displayed in [0:1.5] range (a. u.) and phase images are displayed in [-1:1] range (rad).

Nowadays, deep learning solutions (especially based on convolutional neural networks) are rapidly emerging in optical metrology and imaging[24], including DIHM[25] and lensless microscopy[26]. In DIHM, deep learning found its applications, among others, in classification[27], 3D imaging and localization[28] and autofocusing[29] tasks. Supervised deep learning was also employed in twin-image suppression. The first work with this application was presented by A. Sinha et al.[30], where the convolutional neural network (CNN) was used to directly restore twin-image-free object phase from the input hologram. A different approach was presented by Y. Rivenson et al.[31], where the CNN was trained to remove twin-image from already backpropagated holograms, making this issue less complex. Another solution was presented recently by H. Luo et al.[32], where the network was trained to generate from a single hologram a set of several holograms with different Z distances and then the GS algorithm was applied to obtain twin-image free reconstructions. It is also worth to mention the solutions with unsupervised networks[33–35], which does not need any training dataset. However, their main disadvantage is the need to perform thousands of iterations which makes them extremely time-consuming.

In general, as shown by previous works, supervised neural networks can be effective in twin-image suppression. However, to train a CNN properly, it is needed to be fed with an adequate training dataset. It requires to collect a large number of experimental holograms (which is time-consuming) and most importantly, to assign a ground truth twin-image free reconstruction to each of the experimental holograms. To do that, a significantly more complex setup than DIHM system is usually exploited: A. Sinha et al.[30] and H. Wang et al.[36] used a system with spatial light modulator (SLM), where patterns given at SLM were set as the ground truth phase, Y. Rivenson et al.[31] and H. Luo et al.[32] employed a setup with motorized linear stage to collect several holograms with different Z distances and then used a multi-hologram GS to obtain ground truth reconstruction (thus proposed CNN estimated outcomes of GS algorithm), I. Moon et al.[37] employed an off-axis holography system to obtain ground truth reconstruction and numerically shifted side peaks of the holograms Fourier spectrum to their center to obtain in-line holograms, and Y. Wu et al.[38], Z. Tian et al.[39] and L. Chen et al.[40] used brightfield microscopy images as network target images.

Another issue with supervised CNNs is their universality. When preparing an experimental dataset for network training, it is extremely difficult to achieve high dataset diversity. Therefore, usually, CNNs that are trained on experimental data exhibit low dataset diversity, which makes them work effectively only for samples very similar to the ones present in the training dataset. Moreover, such CNN is then problematic to transfer to different setups between different research groups and usually requires repeating the cumbersome and time-consuming process of preparing the training dataset and performing the CNN training, which still may not guarantee the direct repeatability of the network results. A step forward in overcoming this issue was proposed recently by H. Chen et al.[41,42], where there were presented networks that enable to reconstruct samples not present in the training data. Nevertheless, those networks still required a cumbersome process of experimental data collection and the network performance differed depending on the used training datasets[41].

Finally, previous solutions also suffered a common downside of CNNs – even when CNN is trained correctly, there is still some uncertainty, whether the final result is consistent with reality or is it something "painted" by the CNN, utterly false in its core but real-life-looking. Quantitative self-verification of the method is thus pivotal to achieve its universality.

In this work, we propose a novel, universal twin-image removal network called UTIRnet, which was trained fully on synthetic images, as the physical properties of defocusing (propagating) optical fields are well-known and incorporated via numerical complex optical field propagator. Thanks to that, UTIRnet requires neither the time-consuming process of collecting and labelling experimental data, nor the necessity to upgrade the simple DIHM system with any additional parts needed to obtain ground truth reconstructions. Moreover, because we used only numerically generated images, we were able to train CNN with a large variety of training dataset. Thanks to that, UTIRnet has learned how to generally filter out the twin-image from the backpropagated optical field in a specific system (following the physics of DIHM working principle) instead of learning how to reconstruct specific samples present in the training dataset, which made it universal for practically any kind of samples. The novelty of our solution is showcased also in the incorporation of a single iteration GS algorithm into the processing patch, that made the UTIRnet result to some significant extent physically consistent with the input holograms and therefore with the measured object. We validated the performance of our network both on synthetic and experimental holograms acquired in a lensless DIHM system, showing that it can efficiently suppress the twin-image from all tested samples (amplitude, phase, artificial and biological), even though they were significantly different from data used for network training. This was achieved, for the first time, by physics-based learning oriented on "how to minimize a twin-image phenomenon" and not on "how to retrieve a given object sharply". To enable a wider audience to conveniently use our solution, we released open-source Matlab codes that allow for straightforward and fully automatic generation of UTIRnet for a given set of system parameters[43].

## 2 Methods

*2.1 Network architecture and training*

Figure 2 presents the diagram of the reconstruction process of our novel deep neural network, along with exemplary reconstruction (for hologram shown in Fig. 1) and exemplifying images used for CNNs training. UTIRnet is composed of the angular spectrum (AS)[14,44] propagation blocks and 2 separate CNNs, the first one trained to filter out the twin-image from the amplitude part ($CNN_A$) and the second one trained to filter out the phase part ($CNN_P$) of the optical field. Reconstruction process begins with backpropagating the square root of the hologram intensity, hence its amplitude distribution, to the object plane (at +Z distance). Then, the amplitude and phase parts of the optical field are filtered with $CNN_A$ and $CNN_P$ respectively. After that, the filtered complex optical field is propagated to the hologram plane (at -Z distance), where its amplitude part is replaced with the square root of the recorded hologram. The resulting optical field (replaced amplitude and preserved phase) is then backpropagated to the object plane, where the final twin-image minimized reconstruction is obtained with a single algorithm iteration.

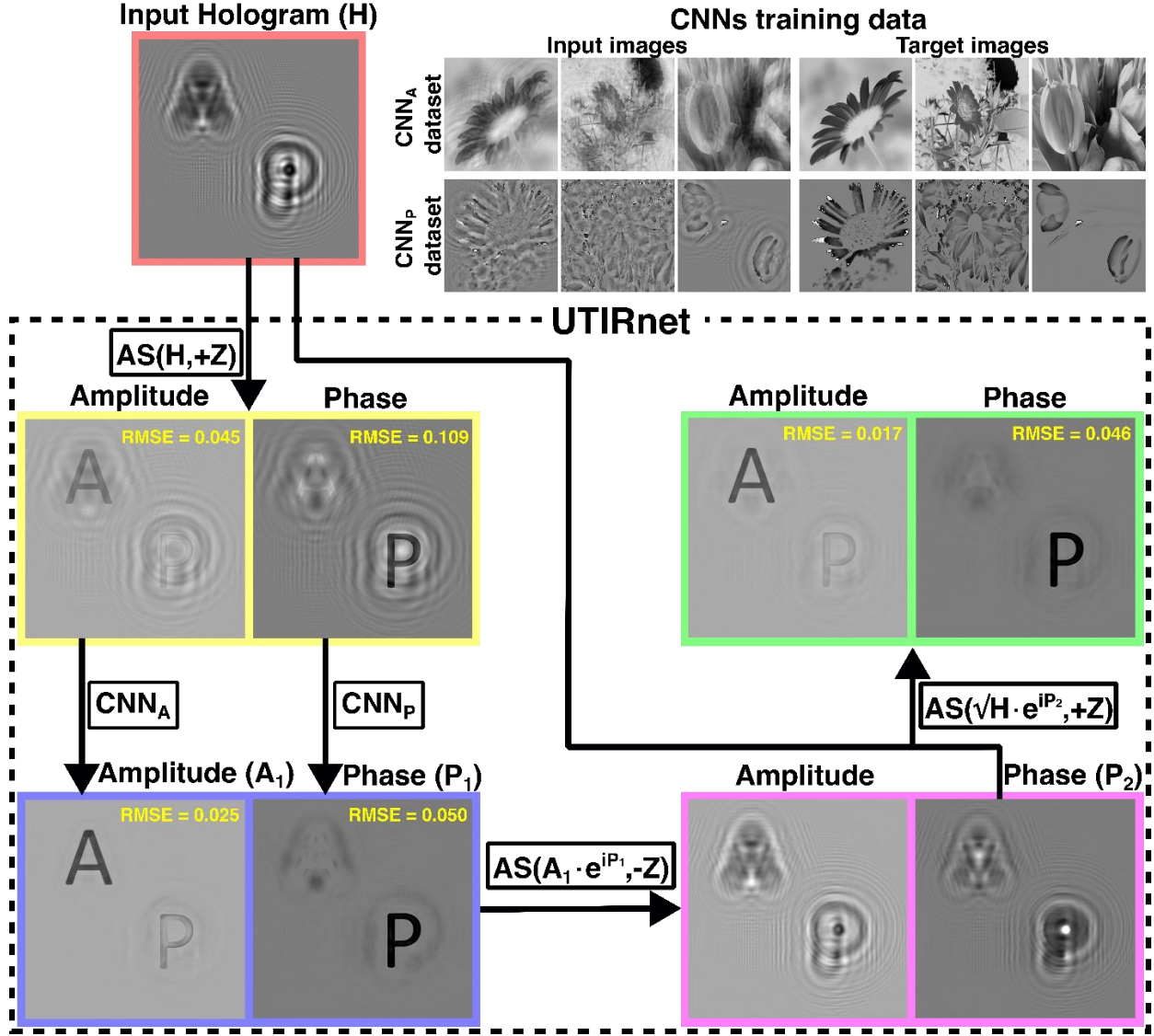

**Fig. 2** Diagram of UTIRnet processing patch along with network exemplary training data. AS(X,Z) – angular spectrum propagation of the X optical field at +Z or -Z distance. $CNN_A$ and $CNN_P$ – convolutional neural networks trained for filtering twin-image from amplitude and phase parts of optical field respectively. Shown root-mean-square errors (RMSE) corroborate the reconstruction improvement after specified operations. Amplitude images are displayed in [0:1.5] range (a. u.) and phase images are displayed in [-1:1] range (rad).

The architecture of employed CNNs, as illustrated in Figure 3, was inspired by the work of S. Feng et al.[45] where the CNN was employed for optical fringe pattern processing. Similar layouts were successfully implemented in numerous works regarding QPI data processing, proving its usefulness in regression tasks[31,36,46,47]. Our network consists of convolutional layers, with each layer comprising 70 filters that perform convolution operations on the data and ReLU activation layers which introduce nonlinearity to the solution. Additionally, the network is divided into two paths, with the second path (bottom path in Figure 3) containing an additional 2x2 px pooling layer to reduce the width and height of the image by a factor of 2 and enable interactions between different pixels. This path concludes with an upsampling layer to restore the original dimensions of the image. The results from both paths are combined in the concentration block and then processed by a final convolutional layer to obtain the network output.

The training process was conducted using a training dataset consisting of 1950 images sized $512 \times 512$ pixels. During training, the mini-batch size was set to 1, and the initial learning rate was equal $10^{-4}$. The learning rate was updated every 5 epochs and reduced by a factor of 5 to help the loss function escape local minima. The ADAM optimizer was employed as the solver for training the network, and the mean-squared-error function served as the loss function. The training process lasted for 30 epochs, which proved sufficient for the networks to converge as no significant further decrease in the loss function was observed thereafter. The networks were trained on a computer equipped with an AMD Ryzen 9 5900X 12-Core 3.70 GHz processor and an NVIDIA GeForce RTX 3080 graphics card with 12 GB of memory, enabling the training of a single network in approximately 500 minutes. It is important to note that this time-consuming training process only needs to be performed once for a given network.

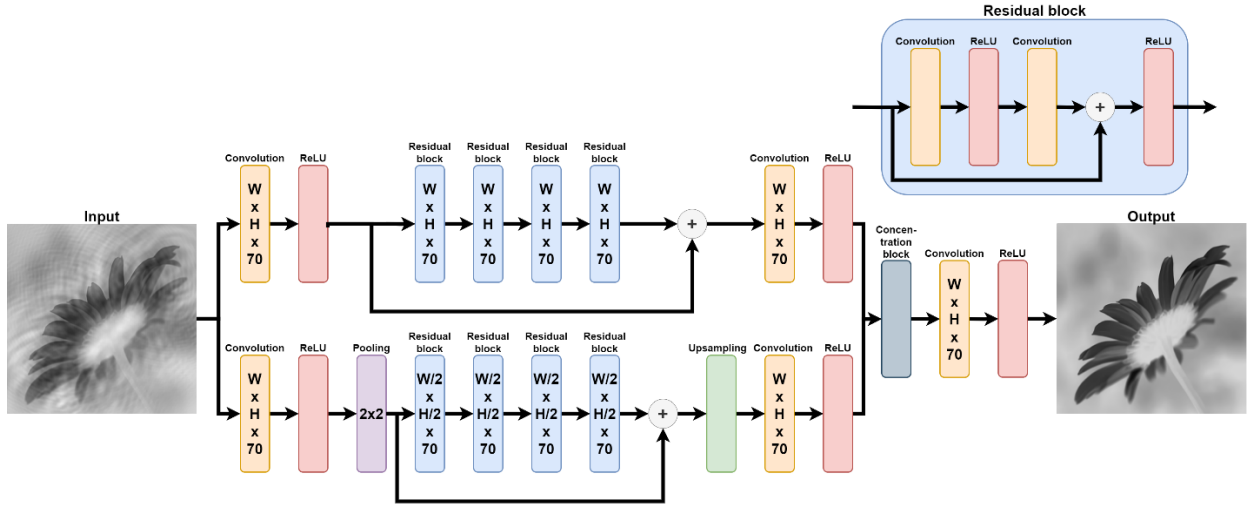

**Fig. 3** The architecture of the used CNNs (CNN$_A$ and CNN$_P$ in Fig. 2).

*2.2 Dataset generation*

The key factor for successful CNNs performance is to provide it with an adequate training dataset. In our case, since we train networks with numerically generated datasets, it is extremely important to generate them as closely as possible to the actual DIHM data registration process. To do that, we were basing on images from the open-source "Flowers recognition" images repository[48]. It is important to note that practically any dataset of well-differentiated pictures (e.g., pictures of different objects) could be suitable for this purpose, and we choose "Flower recognition" mainly because of its easy accessibility and popularity (successfully employed in many deep learning and machine learning applications). Specifically, we utilized 650 images each from the "daisy," "sunflower," and "tulip" sub-datasets of the "Flowers recognition" dataset, resulting in a total of 1950 training images. Validation data were obtained from the "dandelion" and "rose" sub-datasets.

To generate the ground truth target images for network training (representing twin-image-free images that the network aims to achieve), the raw images from the selected repository were first converted to grayscale and resized to 512x512 pixels. They were then denoised using a state-of-the-art block-matching and 3D filtering method[49] to reduce the influence of noise present in the dataset on network training. For CNN$_P$ training, the images underwent an additional high-pass filtering step with a Gaussian kernel of standard deviation equal to 10, in order to remove low-frequency phase information that cannot be recovered by DIHM. Finally, the images were normalized to a range of [0:1] for CNN$_A$ training or randomly within the ranges [-2π:0] and [-π/2:0] for CNN$_P$ training. In the case of CNN$_P$ training, the images were further wrapped to the range of [-π:π] to ensure the presence of 2π phase jumps in the training data (enabling our network to reconstruct those, see Fig. 8) and added with constant +π value (so the final phase range is [0:2π]) to avoid processing negative pixel values by the network.

The network input images, which represent reconstructions affected by the twin-image effect, were generated based on the previously created target images. First, the target images were padded to a size of 1024x1024 pixels (256 pixels in each direction) by repeating the border pixels to avoid aliasing during propagation. Next, objects' optical field were created using the equation $O = A \cdot e^{i \cdot P}$, where *A* represents the padded images and *P* is set to 0 for CNN$_A$ training, or *A* is set to 1 and *P* represents the padded images for CNN$_P$ training. These optical fields were then propagated to the camera plane using the angular spectrum method with the parameters of used DIHM system (including wavelength, effective pixel size, and propagation distance). At the camera plane, the phase components of the optical fields were removed, leaving only the amplitude distributions (square of this distribution is a simulated hologram in camera plane), which were then propagated back to the sample plane. Finally, the backpropagated optical fields were cropped to a size of 512x512 pixels, and their amplitude parts were set as the network input images for CNN$_A$ training, or their phase parts were set as the network input images for CNN$_P$ training.

As it was stated in introduction section, DIHM technique works only with weekly scattering samples and images present in "Flower recognition" dataset, due to their complexity, may seem to not meet this criterion. However, it is to underline that those images were used to simulate a single-plane objects, therefore, there is no multiple scattering phenomenon (single beam of light is scattered only once in object plane). As can be observed in CNNs training data images in Fig. 2, input images differs from training images only by the presence of twin-image effect, proving that weakly scattering assumption is fulfilled.

*2.3 Experimental data reconstruction*

When working with experimental data, the collected holograms need to undergo appropriate processing to ensure proper reconstruction by UTIRnet. Firstly, the optical field of the backpropagated hologram must be normalized to align with the normalization process applied during the network training. This involves normalizing the obtained amplitude by dividing it by its median value and adjusting the obtained phase to the range $[0:2\pi]$ by adding a value of $\pi$. This normalization process is reversed after the CNNs filtering is applied.

Secondly, the CNNs were trained using 512x512 pixel images, while experimental holograms often have larger dimensions. To process larger holograms, we divide the backpropagated optical field into smaller regions of 512x512 pixels, with a 10% overlap between neighboring regions. Each of these regions is then individually processed by the CNNs, and the resulting filtered optical fields are stitched together. Within the 10% overlap areas, the filtered neighboring regions are blended together using an alpha blending technique to achieve a smooth transition between adjacent regions.

It is worthy to notice that the character of the twin-image effect differs, depending on the system parameters (camera-sample distance, light source wavelength, system magnification), therefore, UTIRnet trained for a specific system parameters may not work as well for data collected with significantly different parameters. Nevertheless, when the system parameters do not differ by more than 20-40%, the reconstruction results should be close to optimal (see Supplementary material 1). For the purpose of this article, we were collecting holograms in a lensless holographic microscope system with a camera pixel size of 2.4 μm (ALVIUM Camera 1800 U-2050m mono Bareboard), a diode laser of 405 nm wavelength (CNI Lasers MDL-III-405-20mW) and with camera-light source distance equal 30 cm. In different experiments studied in this paper, we were simulating/collecting holograms for varying camera-sample distances. For data in Figs. 5-6 it was 2.6 mm (which is close to the minimal achievable distance in our system) and for all other Figures it was 17 mm (which is enough to fit a sample in a Petri dish into the system). UTIRnets trained for appropriate system parameters were used in the corresponding experiments.

## 3   Results

In Figure 4, there are shown reconstruction root-mean-square errors (RMSE) for classical AS backpropagation (which is also an input to $CNN_A$ and $CNN_P$ as shown in Fig. 2) and UTIRnet along with several exemplifying reconstructed images for better visualization. RMSE values were calculated for different, numerically generated amplitude-only samples datasets. Training and validation datasets were generated from the popular, open-source "Flowers recognition" images repository[48], while the test datasets were generated from the open-access "Animals-10" repository[50]. Each dataset was composed of 500 pairs of target and input images (target – twin-image-free amplitude at object plane normalized in 0-1 range, input – corresponding object amplitude with twin-image, reconstructed by AS), generated for different flowers/animals species. As can be observed, calculated RMSE values for UTIRnet were similar for all datasets, independently of their type (training, validation or test), showing that the proposed network is overtrained neither for images used for network training, nor for given kind of images (ones present in "Flowers recognition" repository).

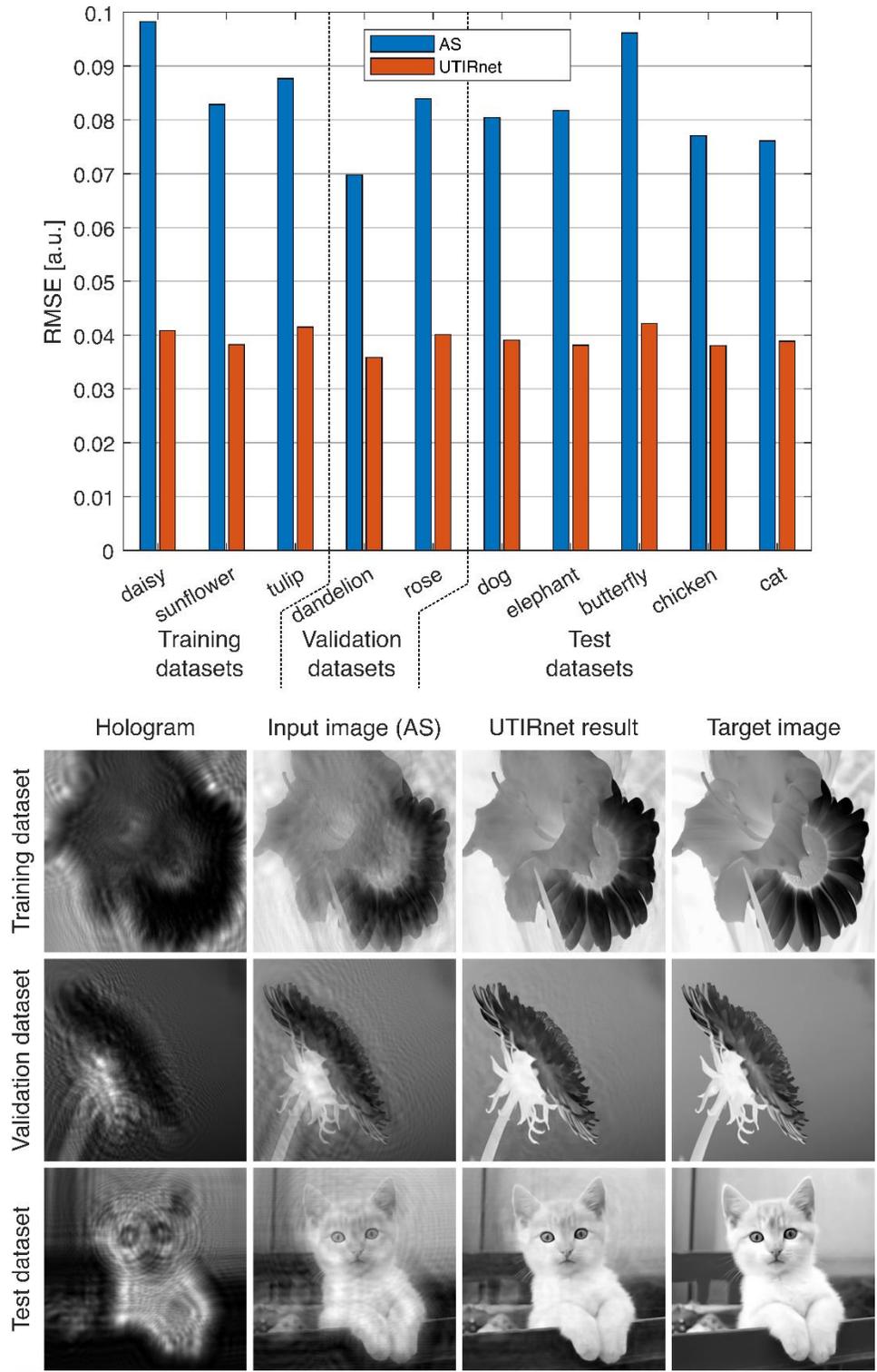

**Fig. 4** Top – RMSE value calculated for 10 different datasets (amplitude-only samples). Bottom – exemplary holograms, amplitude AS and UTIRnet reconstructions and ground truth images from training, validation and test datasets. AS, UTIRnet and target amplitude images are displayed in [0:1.1] range (a. u.).

Figure 5 justifies the incorporation of an additional single iteration GS algorithm (all operations after complex field filtration with CNNs in Fig. 2) into the UTIRnet processing path. Figure 5(a) presents the RMSE values for (1) AS backpropagation, (2) only $CNN_A$ filtration, and (3) the full UTIRnet processing patch for reconstructing the test dataset (composed from all test datasets in Fig. 4 – 2500 images in total). Obtained results show that pure $CNN_A$ filtration reduces the reconstruction error to approximately 65% of AS RMSE value while updating the optical field with the collected hologram in UTIRnet allows to further reduce this error to around 47% of AS error. Figures 5(b)-5(e) shows the exemplary reconstructions of an image from the "rose" dataset in Fig. 4. Original image, Fig. 5(b), is composed of rose flowers in the foreground and a striped chair in the background. In the AS amplitude reconstruction, Fig. 5(c), twin-image is distorting the result. Especially, chairs' striped structure information seems to be lost, as it has a frequency similar to the twin-image fringes. The $CNN_A$, Fig. 5(d), managed to filter out the twin-image from rose flowers correctly, however, with no surprise, it did not manage to recover the chair stripes. On the other hand, UTIRnet, Fig. 5(e), not only retrieved the rose flower correctly, but also managed to partially recover the seemingly lost striped structure information. All these results indicate, that thanks to the deployment of the collected hologram into the reconstruction process, UTIRnet is not only suppressing the twin-image more effectively than the simple CNN filtration but also is able to correct the eventual CNNs errors, making its results to some extent physics-based and reliable (the network does compensate its errors).

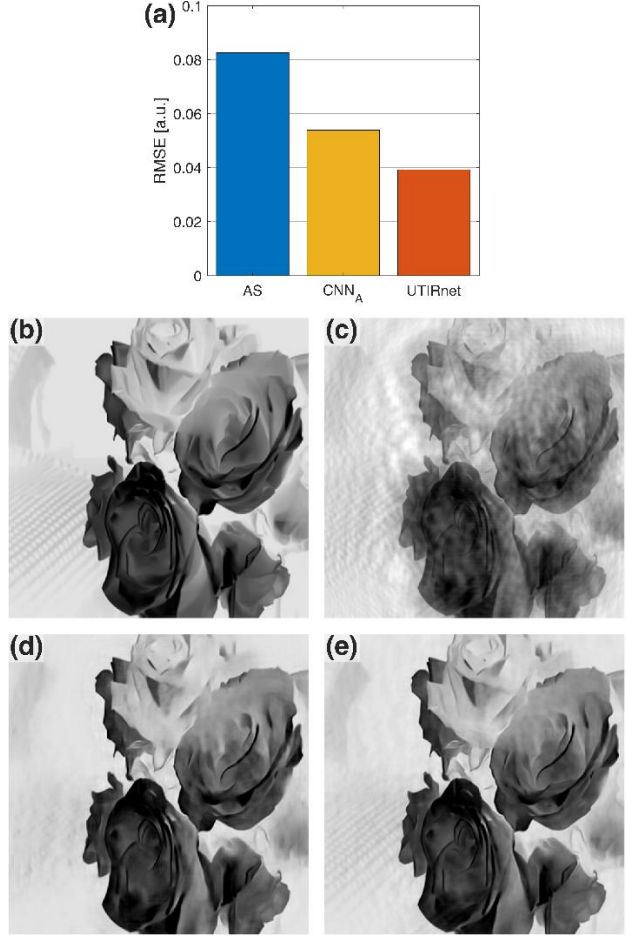

**Fig. 5** (a) RMSE values calculated for the test dataset in Fig. 3 for AS, $CNN_A$ and UTIRnet reconstructions. (b) exemplary target image from "rose" dataset in Fig. 3 and (c)-(e) its AS, $CNN_A$ and UTIRnet reconstructions respectively. Images in (b)-(e) are displayed in [0.2:1.1] range (a. u.).

To demonstrate the effectiveness of the UTIRnet method in minimizing the twin-image for experimental holographic data, we collected a hologram of the USAF amplitude test target. Reconstructed amplitudes obtained using AS backpropagation and UTIRnet are shown in Fig. 6(a) and Fig. 6(b), respectively. We also collected additional hologram with a 561 nm laser (CNI Lasers MGL-FN-561-20mW) and used both holograms (405 nm and 561 nm) to perform reference multi-wavelength GS reconstruction[17] and multi-wavelength GS with additional complex field filtering constraints (GS+CFF) reconstruction[51], as shown in Fig. 6(c) and Fig. 6(d) respectively. In all obtained reconstructions, element 4 from group 7 (line width 2.46 μm) is the smallest distinguishable element, which complies with the system theoretical resolution (2.4 µm – limited by camera pixel size). Comparing UTIRnet with GS, both methods minimized the twin-image similarly, but the neural network produced slightly more accurate results, as GS reconstruction preserved more twin-image artifacts (compare zoomed details in Fig. 6). The GS+CFF algorithm achieved the highest quality reconstruction of the test background due to its nonlinear filtration, which apart from minimizing the twin-image, also reduced the low-frequency background fluctuations. However, when comparing only the presence of twin-image artefacts, UTIRnet and GS+CFF results are similar. As the UTIRnet was not trained to minimize background fluctuations, its reconstruction may be classified as satisfying. On the other hand, when comparing the reconstruction of test elements, GS+CFF sometimes struggled to correctly reconstruct the centers of the elements (low frequency signals), whereas UTIRnet could reconstruct those much more reliably (compare middle horizontal test line in enlarged red area). Importantly, our network achieves these results with the use of only one hologram, demonstrating its ability to work with experimental data directly without the need for additional holograms to be collected (like in the GS-based methods). All this demonstrate that the UTIRnet effectively suppresses the twin-image without compromising the spatial resolution, even though it was trained on a synthetic dataset that did not contain any object similar to the USAF test target.

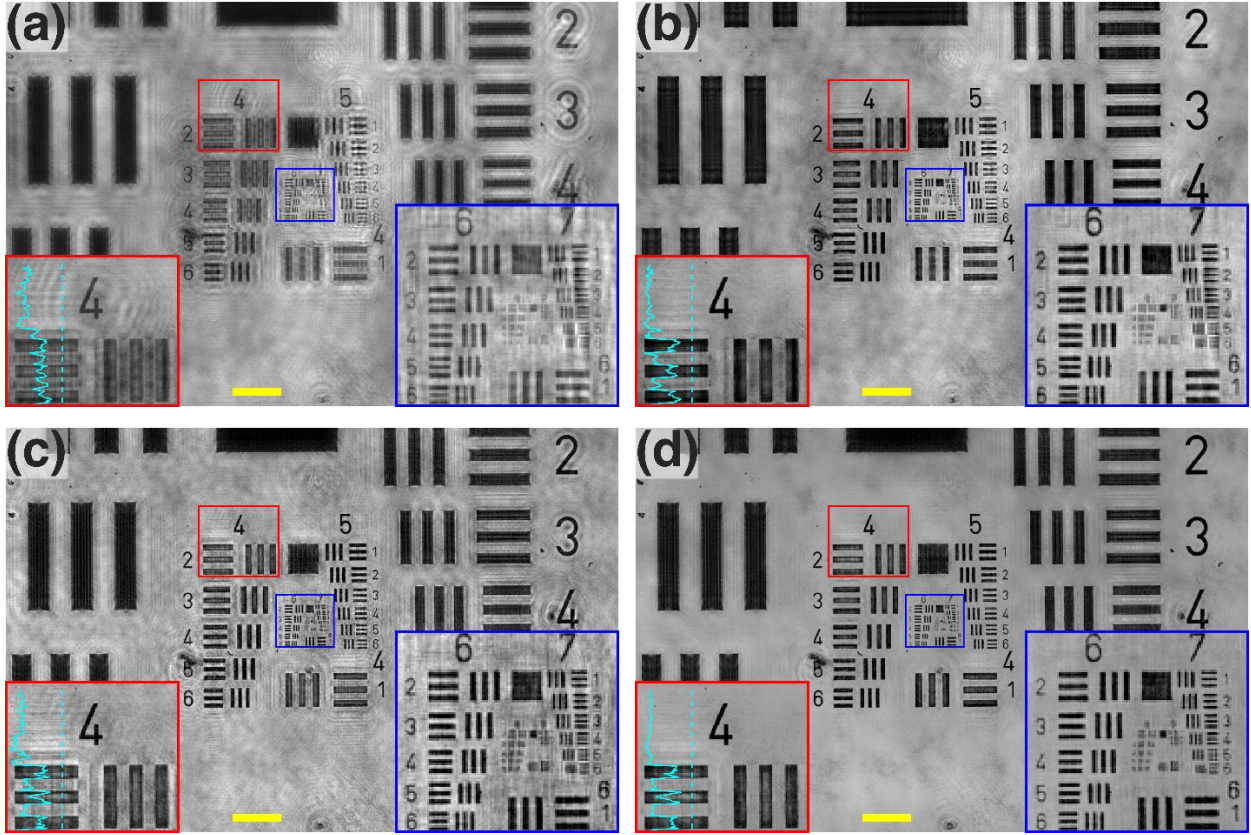

**Fig. 6** Amplitude reconstructions of experimental USAF amplitude test target hologram. (a) AS, (b) UTIRnet, (c) GS, (d) GS+CFF. Yellow scalebars are 200 μm long. All images are displayed in [1:11] range (a. u.).

In Figure 7 the evaluation of UTIRnet in terms of performing quantitative phase measurements is shown. To do that, we collected 2 holograms (first for 405 nm and second for 561 nm wavelength) of the custom-made phase test target (Lyncée Tec, Boroflat 33 glass, 125±5 nm height 15 μm lateral resolution) and reconstructed those holograms with AS, UTIRnet (single hologram reconstruction), GS, and GS+CFF (double hologram reconstruction) methods. As can be observed, compared to the AS method, Fig. 7(a), UTIRnet, Fig. 7(b), successfully minimized the twin-image effect in the reconstructed phase, although, some residual twin-image artifacts remained slightly visible (marked with an orange arrow in Fig. 7(b)). The GS algorithm, Fig. 7(c), also left some remains of the twin-image, visible as dark halos around test elements (known as halo effect[52]). The GS+CFF, Fig. 7(d), seemed to suppress the twin-image from the phase test background the most effectively.

However, when comparing the reconstruction of the square test elements (marked with green arrows in Fig. 7(b)), one can notice that only the UTIRnet managed to recover the phase information in their centers, while the other algorithms significantly lowered the phase values in this area. This omnipresent error resulted from the fact that negative twin-image fringe overlayed the positive phase element, lowering its value and therefore, making the GS and GS+CFF algorithms maintain this lowered phase value. On the other hand, UTIRnet is to some extent aware (basing on the training dataset) of the fact that negative twin-image fringe usually results in lowering phase values, which enables it, as the first method up-to-date, to correct the mentioned error. Figure 7(e) presents the cross-sections through one of the reconstructed elements (sectioned areas are marked with color lines in Figs. 7(a)-7(d)) scaled to nm units. Black, dashed lines mark the sample reference height range (125±5 nm) measured with a white-light interferometer. As can be observed, AS and GS measurements significantly underestimated the element height to around 75 nm, at the same time introducing around -25 nm negative halo values around it, that, if the shape of the test is not known, could be interpreted as dimples next to the test elements. GS+CFF and UTIRned achieved closer, but still slightly underestimated height value equal around 110 nm. However, the GS+CFF measurement tended to overestimate the background height by around 25 nm, therefore, the relative height measured by this method is equal to only 85 nm. To sum up, all tested methods underestimated the calculated height value, although UTIRnet pivotally allowed to significantly minimize this error.

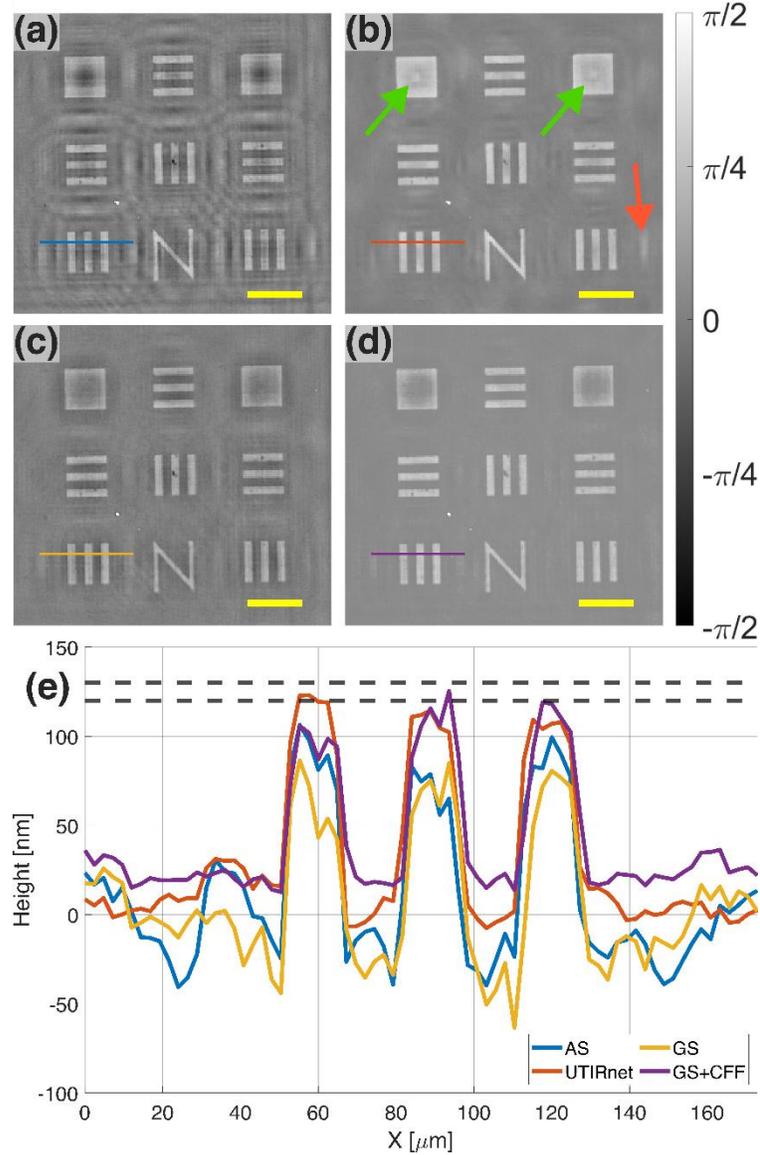

**Fig. 7** Phase reconstructions of experimental custom-made phase test target hologram. (a) AS, (b) UTIRnet, (c) GS, (d) GS+CFF. (e) cross-sections through test element marked with color lines in (a)-(d). Yellow scalebars in (a)-(d) are 100 μm long.

In the subsequent experiment, we investigated the UTIRnet ability to perform quantitative phase imaging of a biological sample (human cheek cells). Cells were collected by scraping the human cheek with a toothpick. Then, the cells were immersed in phosphate buffered saline, placed on a microscope slide (1 mm thick) and covered with cover slip (0.17 mm thick). Figure 8(a) illustrates the full field of view (13.19 x 8.81 mm) cells phase reconstruction using UTIRnet.

Additionally, Figs. 8(b) and 8(c) provide enlarged views of two exemplary regions, reconstructed using the same algorithms as in the previous experiments. Once again, our innovative single-shot network solution successfully eliminated the twin-image, as depicted in Figs. 8(b2) and 8(c2), achieving similar phase quality to GS+CFF, Figs. 8(b4) and 8(c4), while outperforming GS, which exhibited a tendency to distort the reconstructed phase with a halo effect, Figs. 8(b3) and 8(c3). These results also demonstrate that UTIRnet performed well in reconstructing wrapped phase discontinuities, owing to the inclusion of such instances in the training dataset.

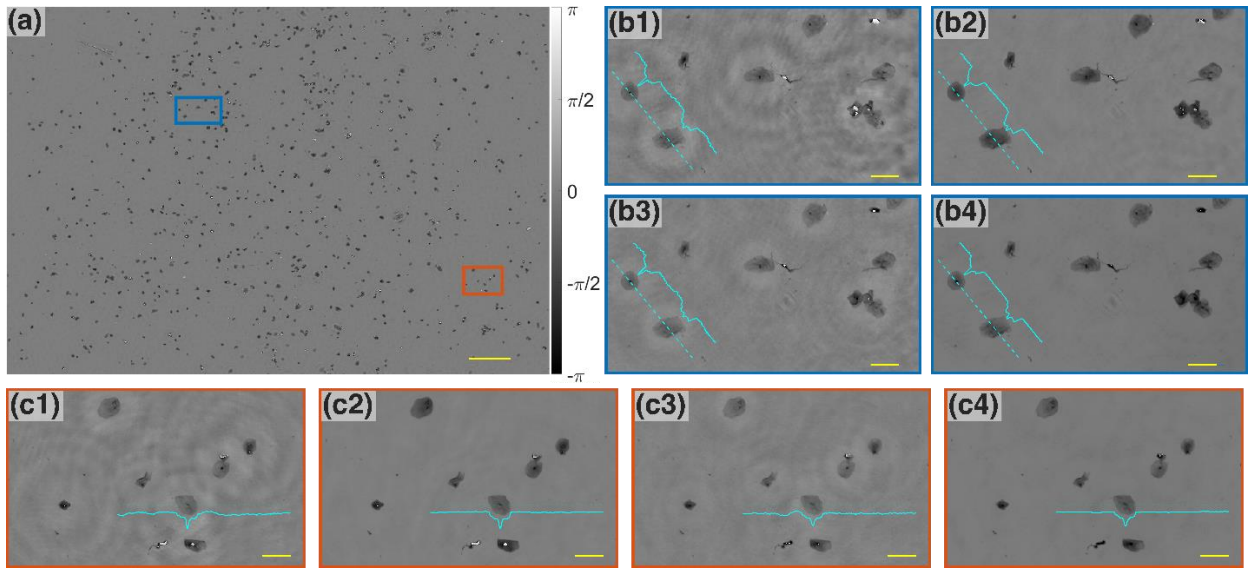

**Fig. 8** Phase reconstructions of experimental human cheek cells hologram. (a) full FOV (13.19 x 8.81 mm) UTIRnet reconstruction, (b1)-(b4) AS, UTIRnet, GS and GS+CFF reconstructions respectively for the area marked with a blue rectangle in (a). (c1)-(c4) AS, UTIRnet, GS and GS+CFF reconstructions respectively for the area marked with an orange rectangle in (a). Yellow scalebar in (a) is 1 mm long and in (b),(c) is 100 μm long.

One of the most intriguing applications of lensless in-line holographic microscopy is its ability to perform live timelapse measurements of dynamic objects with large fields of view. However, achieving high temporal resolution and system stability, free from moving parts that could impact measurement, often requires the use of single-hologram reconstructions over multi-hologram Gerchberg-Saxton-based approaches. Hence, the twin-image is deteriorating both qualitatively the imaging results and quantitatively the diagnostics based on the reconstructed image analysis. In this paper, we demonstrated that UTIRnet can serve as an effective tool for single-hologram, twin-image-free reconstructions, making it a promising candidate for dynamic object measurements provided interframe stability is maintained. To verify this, we performed a timelapse series measurement of live mouse glial restricted progenitors (GRPs) lasting 25 hours, with each subsequent frame captured every 5 minutes. GRPs were isolated from E13 embryos according to a previously described protocol[53]. At 3-4th passage cells were seeded at 30 000 cells/cm2 density on the glass bottom dishes coated with poly-L-lysin and laminin and cultured in standard conditions (37 °C, 5% $CO_2$ concentration) until reaching 50% confluency.

Figure 9 illustrates the reconstructed phases using AS, Fig. 9(a), and UTIRnet, Fig. 9(b), for the first frame, after 1 hour, and after 2 hours. The same cells are marked with color arrows across different frames. Supplementary videos accompanying this article: Video 1 (MOV, 5.1 MB) and Video 2 (MOV, 3 MB) display the full 25-hour timelapse measurements using AS and UTIRnet, respectively, with Video 3 (MOV, 4.1 MB) showing a combined view of both methods for comparison purposes. Extended description of Videos 1-3 is present in Supplementary material 1. Our results demonstrate that UTIRnet effectively suppressed the twin-image and maintained interframe stability, with no visible phase fluctuations or single-frame artifacts, validating its potential for dynamic object measurements.

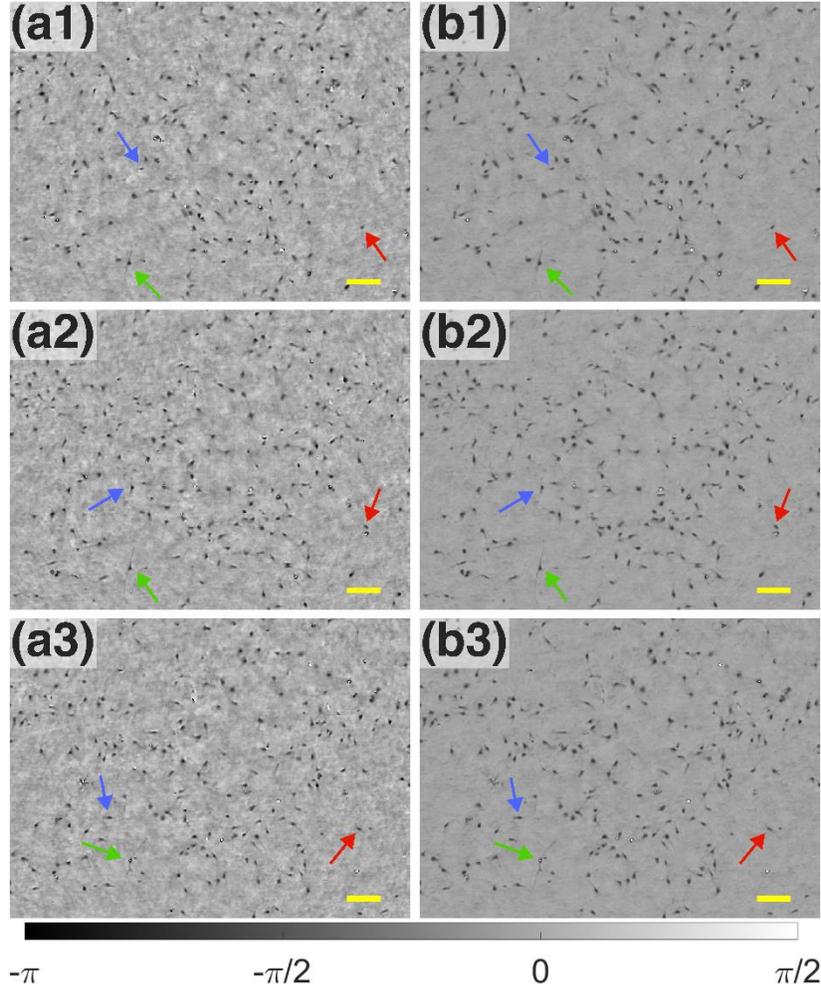

**Fig. 9** Phase reconstructions of experimental GRPs timelapse holograms with (a) AS and (b) UTIRnet methods after (top) 0 h, (middle) 1 h and (bottom) 2 h. Color arrows mark the same cells in different frames. Yellow scalebars are 100 μm long.

Table 1 provides a comparison of the reconstruction times for different algorithms discussed in this article. As can be observed, UTIRnet exhibits slower processing times compared to the second slowest GS+CFF method by an order of magnitude, and compared to the simple AS backpropagation method by two orders of magnitude. This long processing time is a result of the high complexity of UTIRnet's architecture. However, even with this network complexity, the processing time of UTIRnet remains below 1 minute, even for extremely large 4096x4096 pixel images, which should be acceptable for most applications. Furthermore, our network still demonstrates significantly faster processing times compared to single-frame solutions based on regularization (e.g., compressive sensing method is reported to last 35 s for 500x500 px images[22]) or unsupervised learning (e.g., F. Wang et al. report that their network reconstructs 256x256 px images in ~10 min[33]) approaches.

Table 1 Reconstruction times for different algorithms for various image sizes.

| Image size [px] | AS [s] | UTIRnet [s] | GS (5 iter.) [s] | GS+CFF (5 iter.) [s] |
| --- | --- | --- | --- | --- |
| 512x512 | 0.0074 | 0.91 | 0.026 | 0.073 |
| 1024x1024 | 0.018 | 3.93 | 0.075 | 0.16 |
| 2048x2048 | 0.073 | 11.03 | 0.30 | 0.68 |
| 4096x4096 | 0.32 | 36.25 | 1.27 | 5.57 |

## 4 Discussion

In this manuscript we have presented a step forward in deep learning approaches for twin image suppression by proposing a supervised universal neural network called UTIRnet, that is capable of effectively suppressing the twin-image effect in digital in-line holography. The main novelty of UTIRnet is that it is trained solely on numerically generated data, therefore, contrary to networks trained on experimental data, it does not require any time-consuming and complicated process of collecting and labeling training data. The results presented in this paper demonstrate that UTIRnet significantly mitigates the twin-image issue, even when measuring objects that are fundamentally different from those in the training dataset. Another novelty of our solution is the fact that UTIRnet updates the complex optical field filtered by CNNs with input holograms during the reconstruction process, ensuring that the final result is consistent with the actual registered hologram and is therefore in core physics-based, allowing for partial compensation of errors introduced by CNN filtering.

Our proposed network was validated in various experiments performed in a lensless DIHM system and shown to effectively reduce twin-images in both simulated and experimental data. Among these experiments, we showed that UTIRnet successfully reconstructs both amplitude and phase samples (either artificial or biological) with results comparable to the reference multi-frame Gerchberg-Saxton-based methods. Furthermore, performed experiments revealed that UTIRnet does not exhibit any resolution loss, the obtained phase measurements are quantitative and UTIRnet maintains inter-frame stability when performing timelapse series reconstructions.

Despite its many advantages, UTIRnet does have a few issues that must be considered. Firstly, UTIRnet is universal in terms of reconstructing various types of objects but only works in systems with parameters similar to those used for network training (i.e., defocus distance, wavelength, pixel size, or, if used, microscope objective magnification). Training the network to measure samples in DIHM systems with significantly different parameters requires to repeat a time-consuming (but still a one-time event) process of network training. Some solution to this problem may be to generate training dataset with varied system parameters, however, this may negatively affect the network effectivity.

Secondly, despite significant twin-image suppression, it is not completely removed and its residuals are still visible in the final reconstruction. These residuals may be further minimized by increasing CNNs architecture's complexity, however, this would increase both training and reconstruction time. Alternatively, reducing the network's universality and training it on a more specific, numerical dataset may be a solution (e.g., when the network is designed to reconstruct small cells, then the training dataset may be narrowed down to only small objects).

Finally, UTIRnet is capable of performing real-time reconstruction (below 1 second) only for images up to 512x512 pixels in size. However, this time increases significantly for larger images (even up to 36 s for 4096x4096 px images), which may be unsatisfying when reconstructing large volumes of data. To speed up the reconstruction process at the expense of longer training times, the network can be trained for larger image sizes which should reduce the number of performed operations. Another solution may be to train CNNs with reduced architecture's complexity. However, this may again reduce network performance in terms of twin-image minimization. Nevertheless, all mentioned UTIRnet drawbacks are rather minor and, if necessary, can be mitigated by modifying the network training process.

## 5 Conclusions

In conclusion, our proposed neural network solution effectively and robustly suppresses the twin-image effect in digital in-line holographic microscopy. It has been validated numerically and on experimental samples (both artificial and biological) collected in a lensless DIHM system. The solution is not limited to our system and can be easily applied to other DIHM configurations using the open-source codes we released with the manuscript. We believe that UTIRnet will advance the development of digital in-line holographic microscopy and allow the wider community to benefit from simple DIHM imaging without a twin-image effect.

**Disclosures**

The authors declare no conflicts of interests.

**Code, Data and Materials Availability**

Data used in manuscript Figs. 5-8 and supplementary material Figs. S1, S3 and S4 are available at Ref.[54]. UTIRnet Matlab codes are available at Ref.[43]. All other data are available from the corresponding author upon reasonable request.

**Funding Sources**

This research has been funded by National Science Center, Poland (2020/39/D/ST7/03236) and by the Grant PID2020-120056GB-C21 funded by MCIN/AEI/10.13039/501100011033. M.R. is supported by the Foundation for Polish Science (FNP start program).

field on-chip microscopy," Nat. Methods **9**(9), 889–895 (2012) [doi:10.1038/nmeth.2114].

10. Y. Wu and A. Ozcan, "Lensless digital holographic microscopy and its applications in biomedicine and environmental monitoring," Methods **136**, 4–16, Elsevier Inc. (2018) [doi:10.1016/j.ymeth.2017.08.013].

11. M. J. Lopera and C. Trujillo, "Holographic point source for digital lensless holographic microscopy," Opt. Lett. **47**(11), 2862, Optica Publishing Group (2022) [doi:10.1364/OL.459146].

12. M. Molaei and J. Sheng, "Imaging bacterial 3D motion using digital in-line holographic microscopy and correlation-based de-noising algorithm," Opt. Express **22**(26), 32119 (2014) [doi:10.1364/OE.22.032119].

13. V. Micó, K. Trindade, and J. Á. Picazo-Bueno, "Phase imaging microscopy under the Gabor regime in a minimally modified regular bright-field microscope," Opt. Express **29**(26), 42738, Optica Publishing Group (2021) [doi:10.1364/OE.444884].

14. T. Latychevskaia and H.-W. Fink, "Practical algorithms for simulation and reconstruction of digital in-line holograms," Appl. Opt. **54**(9), 2424, Optica Publishing Group (2015) [doi:10.1364/AO.54.002424].

15. R. W. Gerchberg, "A practical algorithm for the determination of phase from image and diffraction plane pictures," Optik (Stuttg). **35**, 237–246 (1972).

16. J. R. Fienup, "Phase retrieval algorithms: a comparison," Appl. Opt. **21**(15), 2758 (1982) [doi:10.1364/AO.21.002758].

17. C. Zuo et al., "Lensless phase microscopy and diffraction tomography with multi-angle and multi-wavelength illuminations using a LED matrix," Opt. Express **23**(11), 14314, Optica Publishing Group (2015) [doi:10.1364/OE.23.014314].

18. M. Sanz et al., "Compact, cost-effective and field-portable microscope prototype based on MISHELF microscopy," Sci. Rep. **7**(1), 43291, Nature Publishing Group (2017) [doi:10.1038/srep43291].

19. A. Greenbaum, U. Sikora, and A. Ozcan, "Field-portable wide-field microscopy of dense samples
28

**Caption List**

**Fig. 1** Scheme of a typical lensless DIHM system, simulation of optical fields at different planes and exemplary reconstruction of the hologram with the use of numerical backpropagation. Amplitude images are displayed in [0:1.5] range (a. u.) and phase images are displayed in [-1:1] range (rad).

**Fig. 2** Diagram of UTIRnet processing patch along with network exemplary training data. AS(X,Z) – angular spectrum propagation of the X optical field at +Z or -Z distance. CNNA and CNNP – convolutional neural networks trained for filtering twin-image from amplitude and phase parts of optical field respectively. Shown root-mean-square errors (RMSE) corroborate the reconstruction improvement after specified operations. Amplitude images are displayed in [0:1.5] range (a. u.) and phase images are displayed in [-1:1] range (rad).

**Fig. 3** The architecture of the used CNNs (CNNA and CNNP in Fig. 2).

**Fig. 4** Top – RMSE value calculated for 10 different datasets (amplitude-only samples). Bottom – exemplary holograms, amplitude AS and UTIRnet reconstructions and ground truth images from training, validation and test datasets. AS, UTIRnet and target amplitude images are displayed in [0:1.1] range (a. u.).

**Fig. 5** (a) RMSE values calculated for the test dataset in Fig. 3 for AS, CNNA and UTIRnet reconstructions. (b) exemplary target image from "rose" dataset in Fig. 3 and (c)-(e) its AS, CNNA and UTIRnet reconstructions respectively. Images in (b)-(e) are displayed in [0.2:1.1] range (a. u.).

**Fig. 6** Amplitude reconstructions of experimental USAF amplitude test target hologram. (a) AS, (b) UTIRnet, (c) GS, (d) GS+CFF. Yellow scalebars are 200 µm long. All images are displayed in [1:11] range (a. u.).

**Fig. 7** Phase reconstructions of experimental custom-made phase test target hologram. (a) AS, (b) UTIRnet, (c) GS, (d) GS+CFF. (e) cross-sections through test element marked with color lines in (a)-(d). Yellow scalebars in (a)-(d) are 100 µm long.

**Fig. 8** Phase reconstructions of experimental human cheek cells hologram. (a) full FOV (13.19 x 8.81 mm) UTIRnet reconstruction, (b1)-(b4) AS, UTIRnet, GS and GS+CFF reconstructions respectively for the area marked with a blue rectangle in (a). (c1)-(c4) AS, UTIRnet, GS and GS+CFF reconstructions respectively for the area marked with an orange rectangle in (a). Yellow scalebar in (a) is 1 mm long and in (b),(c) is 100 µm long.

**Fig. 9** Phase reconstructions of experimental GRPs timelapse holograms with (a) AS and (b) UTIRnet methods after (top) 0 h, (middle) 1 h and (bottom) 2 h. Color arrows mark the same cells in different frames. Yellow scalebars are 100 µm long.

**Table 1** Reconstruction times for different algorithms for various image sizes.

# Supplementary material 1

## UTIRnet performance depending on parameters used for network training

UTIRnet is trained using specific system parameters, including camera pixel size, light source wavelength, camera-sample distance (Z), and system magnification. It should be noted that the characteristics of the twin-image artifact differ depending on these parameters (see images in Fig. S1). Consequently, when UTIRnet is utilized to reconstruct holograms acquired with different system parameters than those used during network training, its performance may be compromised. Figure S2(a) presents plots of the reconstruction root-mean-square error (RMSE) for angular spectrum (AS) backpropagation and two UTIRnets employed in this manuscript. The first network was trained with a camera-sample distance of $Z_1$ = 2.6 mm, while the second network was trained with $Z_2$ = 17 mm. The RMSE values were calculated for simulated datasets generated with various camera-sample distances ranging from 1 to 25 mm (in steps of 0.8 mm). Each dataset consisted of 2500 images, where the target ground truth images were the same as those used in the test datasets presented in manuscript Fig. 3, and the input images were generated for the corresponding Z value. Figure S2(b) displays the corresponding relative RMSE values, calculated as $\frac{\text{UTIRnet RMSE}}{\text{AS RMSE}}$ 100%. As anticipated, the lowest relative RMSE values were obtained when the camera-sample distance matched the training distance. However, even significant changes in system parameters – up to 20% (±0.5 mm) for Z = 2.6 mm or 40% (±7 mm) for Z = 17 mm – resulted in a relative RMSE increase of no more than 2%. This finding demonstrates that there is no necessity to retrain UTIRnet after every insignificant modification in the experimental system setup.

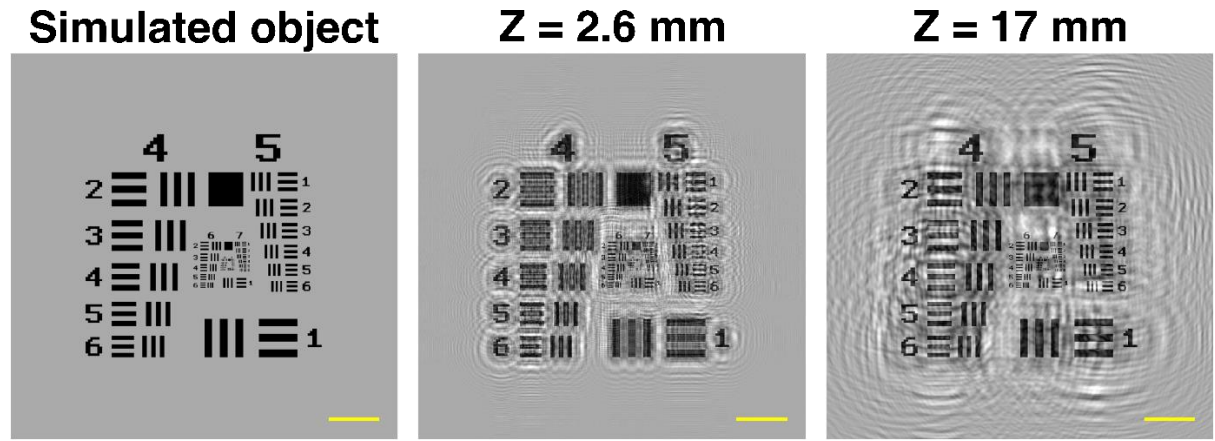

*Figure S10. Simulated amplitude object and reconstructed holograms with AS backpropagation for camera-sample distances equal 2.6 and 17 mm. This example shows the difference in twin-image character for different camera-sample distances. Yellow scalebar is 200 µm long.*

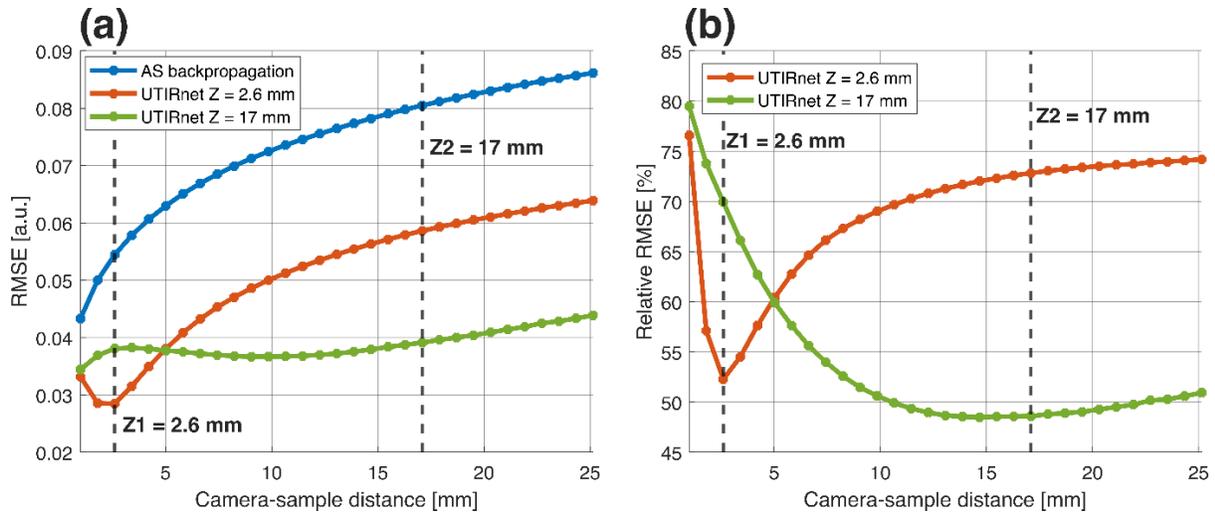

*Figure S11. RMSE values (a) and relative RMSE values in relation to AS method (b) for reconstructing synthetic dataset generated with different camera-sample distances. Results are shown for AS backpropagation and two UTIRnets trained for different camera-sample distances (Z1 = 2.6 mm and Z2 = 17 mm).*

## Supplementary Videos information

Supplementary Videos 1-3 present a timelapse reconstruction result of live glial restricted progenitors lasting 25 hours, with each subsequent frame captured every 5 minutes. The reconstructed full field of view is 5496 x 3672 px (13.19 x 8.81 mm) size and is presented in Fig. S3. In Videos 1-3 there are presented the 480 x 360 px (1.15 x 0.86 mm) region of interests marked with red rectangle in Fig. S3.

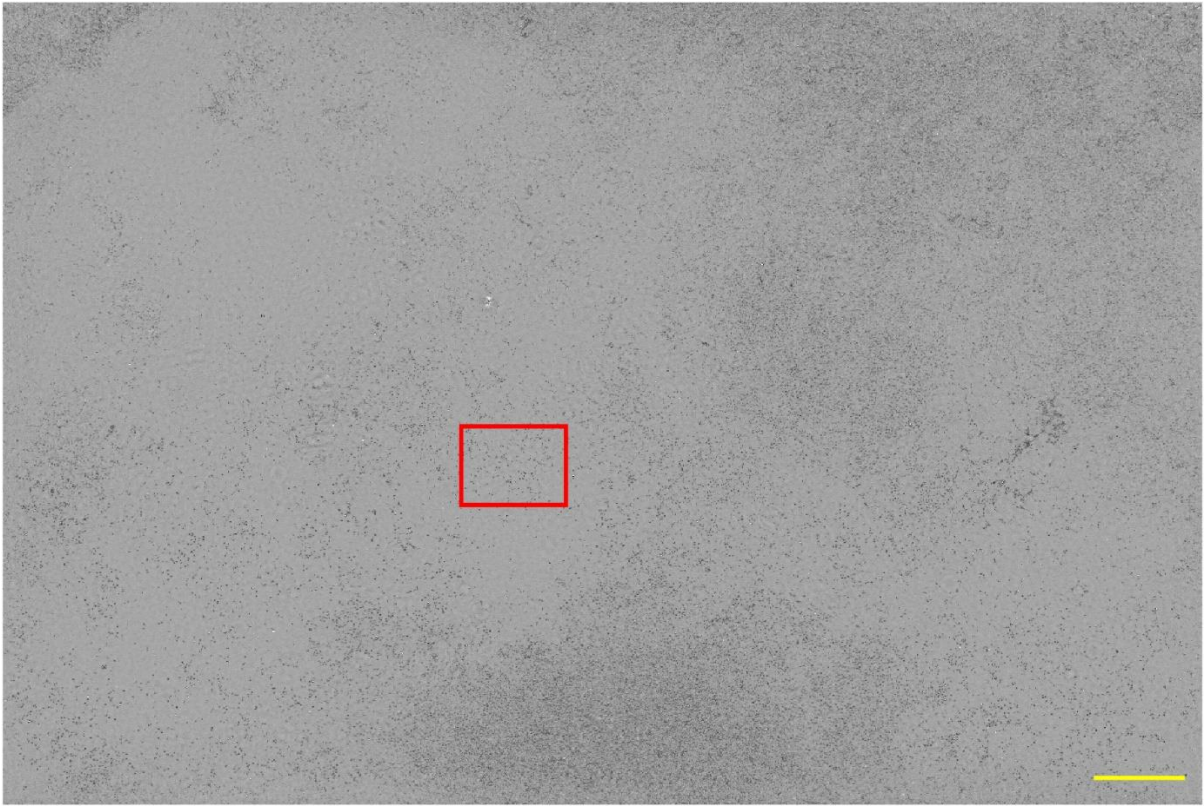

*Figure S12. Full field of view glial restricted progenitors UTIRnet phase reconstruction. Red rectangle marks the region of interest shown in Videos S1-S3. Yellow scalebar is 1 mm long.*

Video 1 presents the reconstruction obtained with AS method (spoiled with twin-image), Video 2 presents the UTIRnet reconstruction (with twin-image minimized), whereas Video 3 shows a combination of both reconstructions (video left side – AS, video right side – UTIRnet) for comparison purposes. The first frames of each video are shown in Fig. S4.

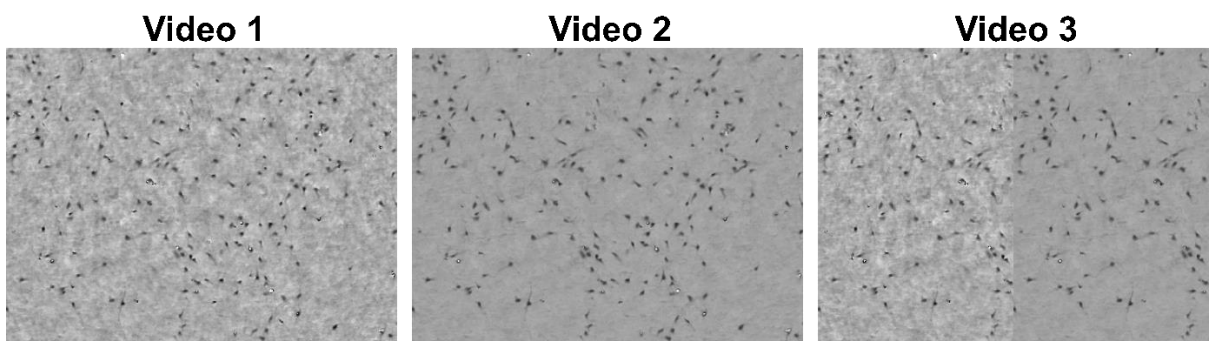

*Figure S13. First frames of videos 1-3. Video 1 – AS reconstruction; Video 2 – UTIRnet reconstruction; Video 3 – AS reconstruction is shown in the left half and UTIRnet reconstruction is shown in the right half of the video.*

Videos 1-3 may be downloaded at: http://gofile.me/67C1K/2ScnyllW9